\documentclass[journal = nalefd, manuscript = letter, layout = twocolumn]{achemso}

\usepackage[T1]{fontenc} 
\usepackage{units}
\usepackage{amssymb}

\author{Simon Geyer}
\affiliation{Department of Physics, University of Basel, Klingelbergstrasse 82, CH-4056 Basel, Switzerland}

\author{Leon C. Camenzind}
\affiliation{Department of Physics, University of Basel, Klingelbergstrasse 82, CH-4056 Basel, Switzerland}

\author{Lukas Czornomaz}
\affiliation{IBM Research-Z\"urich, S\"aumerstrasse 4, CH-8803 R\"uschlikon, Switzerland}

\author{Veeresh Deshpande}
\affiliation{IBM Research-Z\"urich, S\"aumerstrasse 4, CH-8803 R\"uschlikon, Switzerland}
\altaffiliation
{Current address: Institute IFOX, Helmholtz Zentrum Berlin f\"ur Materialien und Energie, Hahn-Meitner Platz 1, D-14109 Berlin, Germany }

\author{Andreas Fuhrer}
\affiliation{IBM Research-Z\"urich, S\"aumerstrasse 4, CH-8803 R\"uschlikon, Switzerland}

\author{Richard J. Warburton}
\affiliation{Department of Physics, University of Basel, Klingelbergstrasse 82, CH-4056 Basel, Switzerland}

\author{Dominik M. Zumb\"uhl}
\affiliation{Department of Physics, University of Basel, Klingelbergstrasse 82, CH-4056 Basel, Switzerland}
\email{dominik.zumbuhl@unibas.ch}

\author{Andreas V. Kuhlmann}
\affiliation{Department of Physics, University of Basel, Klingelbergstrasse 82, CH-4056 Basel, Switzerland}
\alsoaffiliation{IBM Research-Z\"urich, S\"aumerstrasse 4, CH-8803 R\"uschlikon, Switzerland}
\email{andreas.kuhlmann@unibas.ch}

\title{Silicon quantum dot devices with a self-aligned second gate layer}


\begin{document}

\begin{abstract}
We implement silicon quantum dot devices with two layers of gate electrodes using a self-alignment technique, which allows for ultra-small gate lengths and intrinsically perfect layer-to-layer alignment. In a double quantum dot system, we investigate hole transport and observe current rectification due to Pauli spin blockade. Magnetic field measurements indicate that hole spin relaxation is dominated by spin-orbit interaction, and enable us to determine the effective hole $g$-factor $\simeq1.6$. From an avoided singlet-triplet crossing, occurring at high magnetic field, the spin-orbit coupling strength $\unit[\simeq0.27]{meV}$ is obtained, promising fast and all-electrical spin control. 
\end{abstract}
\vspace{0.5cm}

Classical silicon (Si) integrated circuits, hosting billions of metal-oxide-semiconductor field-effect transistors (MOSFET), are the prototypical example for scalable electronic platforms. Si MOS quantum dots\citep{Angus2007,Lim2009} represent the quantum analogon to a classical MOSFET and, as  hosts of coherent and high fidelity spin qubits\cite{Loss1998,Kloeffel2013,Veldhorst2014,Petit2020,Yang2020,Yoneda2017,Zajac2017,Watson2018,Maurand2016}, are prime candidates for scaling up quantum circuits\cite{Veldhorst2017,Vandersypen2017}. While a basic MOSFET is a three-terminal device, a multi-layer gate stack is typically employed for quantum dot formation and qubit control\cite{Angus2007,Lim2009,Veldhorst2014,Petit2020,Yang2020,Eenink2019,Lawrie2020}. In addition, fabrication is demanding due to tight requirements on feature size and layer-to-layer alignment accuracy. 

Device integration can be facilitated by introducing a self-alignment technique. Self-aligned processes have been successfully applied in electronics industry for many years. The basic idea is to use an existing patterned structure on a device to define the pattern of a subsequent process. A prime example is to use the gate of a MOSFET as a mask for defining the source and drain regions by means of ion implantation\cite{Vadasz1969,Bower1968}.

\begin{figure*}
\centering \includegraphics[width=\textwidth]{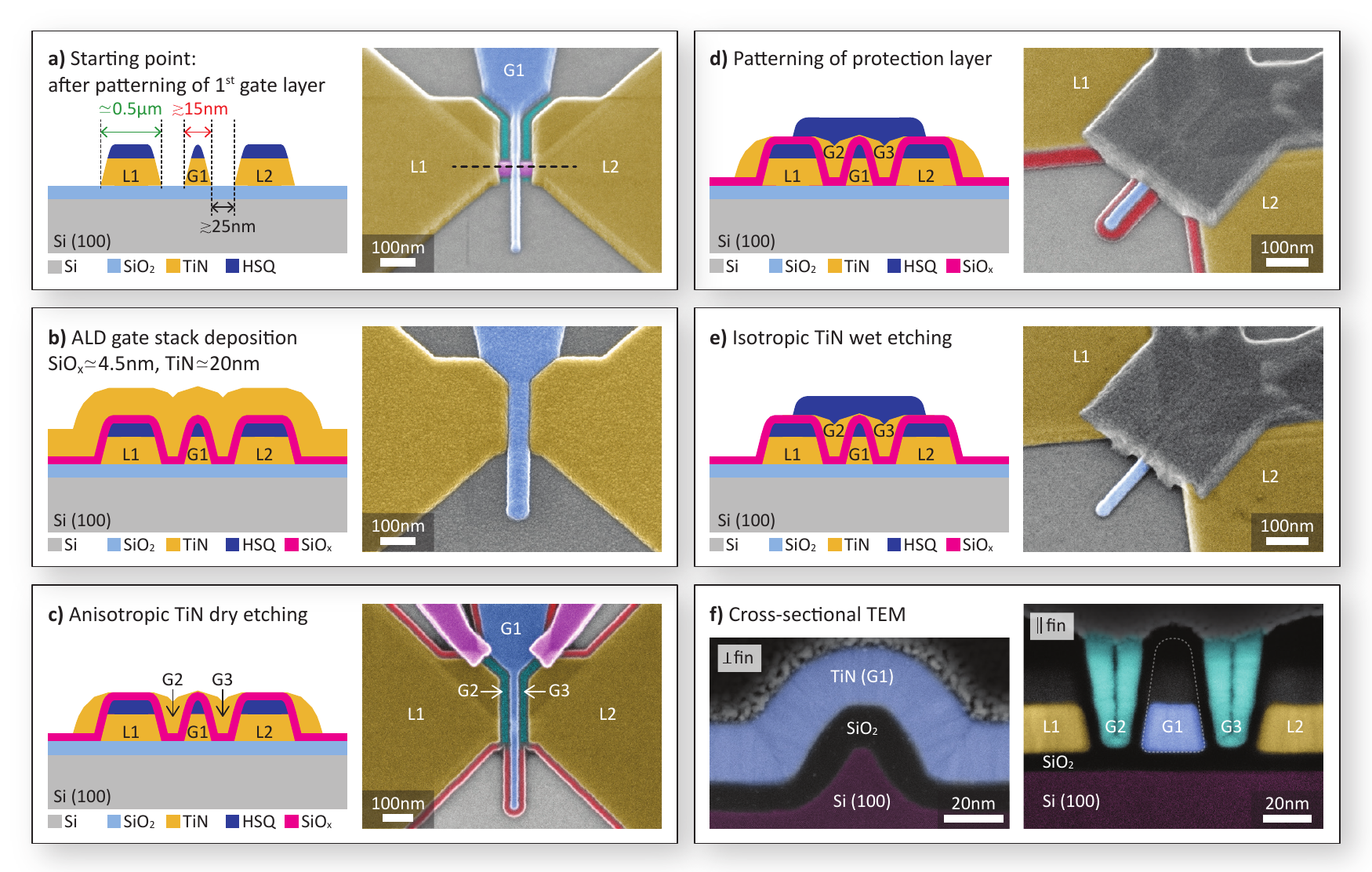}
\caption{\textbf{Fabrication process flow}. (a-e, left panel) Schematic device cross-section along the black dashed line of the (a) SEM image. The horizontal axes of the left and right panels are scaled differently. (a) Device with completed first gate layer, consisting of two lead gates L1\,\&\,L2 (yellow in SEM image) and a central finger gate G1 (blue) that is wrapped around the Si fin (magenta). EBL with hydrogen silsesquioxane (HSQ) resist is employed for gate definition. The gaps separating the gates (turquoise) act as a template for the second gate layer. (b) Deposition of the gate stack by means of ALD results in a uniform surface coverage, such that the gaps are almost evenly filled with material. (c) TiN is removed from the flat surfaces, which are not protected by resist (magenta), by timed dry etching. TiN residues (red) at topography steps still connect gates G2\,\&\,G3 of the second gate layer. (d, e) A protective resist mask is applied to remove all unintended TiN residues with an isotropic wet etch. (f) Cross-sectional TEM images perpendicular (left panel) and parallel (right panel) to the fin. Left: the quantum dot is induced at the apex of the roughly triangular-shaped Si fin (purple). Right: Gates G2\,\&\,G3 (turquoise) are perfectly aligned relative to the first gate layer. Good electrical isolation is ensured by a thin SiO$_{x}$ layer sandwiched between the two gate layers.}
\label{fig1}
\end{figure*}

While most experiments employ electron spin states\cite{Veldhorst2014,Petit2020,Yang2020,Yoneda2017,Zajac2017,Watson2018} to encode quantum information, hole spin qubits\cite{Maurand2016,Watzinger2018,Hendrickx2020} represent an attractive alternative, particularly for large-scale quantum circuits. Hole spins in Si experience a strong spin-orbit interaction (SOI), enabling rapid and all-electrical spin control\cite{Maurand2016,Voisin2015}. Electron spin manipulation, in contrast, is driven by magnetic fields, requiring additional device components such as a micromagnet\cite{PioroLadriere2008,Yang2020} or transmission line\cite{Veldhorst2014,Petit2020}, which hamper scalability. Furthermore, an exceptionally strong and electrically controllable SOI has been predicted for holes in Si nanowires\cite{Kloeffel2018,Froning2020}. 

In this work, Si fin-field-effect-transistor-like devices\cite{Maurand2016,Kuhlmann2018} with two layers of gates are used to form hole double dots. The second gate layer is realized using self-alignment, where the first gate layer acts as a mask for the subsequent one. With this approach, multi-gate stack integration is facilitated, and ultra-small gate lengths $\unit[\simeq15]{nm}$ as well as perfect layer-to-layer alignment can be achieved. We study charge transport for holes in the Pauli spin blockade (PSB) regime and identify the SOI as the dominant mechanism for lifting spin blockade. Both the SOI coupling strength and the effective hole $g$-factor are obtained from an anticrossing between singlet and triplet spin states at high magnetic field.

\begin{figure*}
\centering\includegraphics[width=\textwidth]{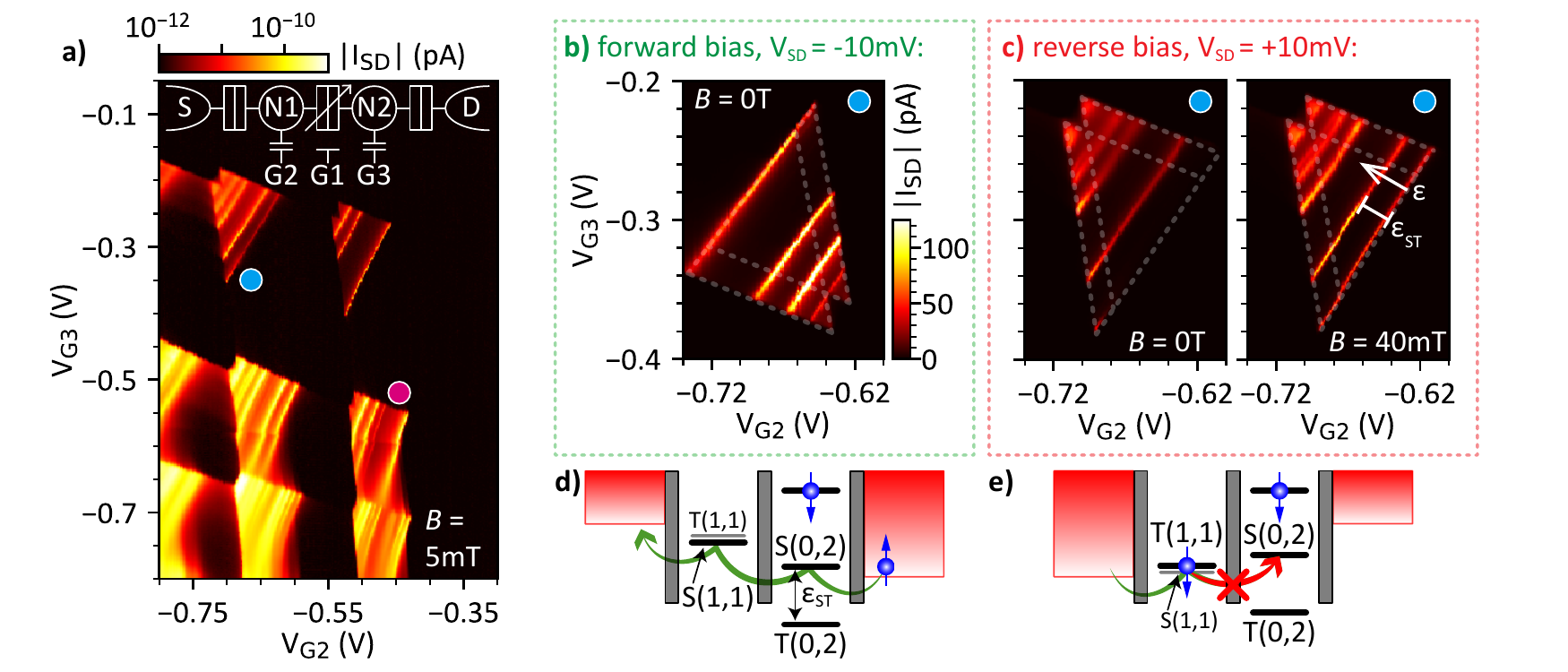}
\caption{\textbf{Bias triangles and Pauli spin blockade for holes}. (a) Double dot charge stability diagram measured for V$_{\mathrm{SD}}=\unit[+10]{mV}$ and V$_{\mathrm{G1}}=\unit[-770]{mV}$. The simplified equivalent circuit of the device is depicted in the inset. While the dot occupancies N1 and N2 are separately controlled by gates G2 and G3, the inter-dot tunnel barrier is tuned by G1. The colored circles mark the pairs of bias triangles for which signatures of PSB are observed. A close-up of the triangles indicated by the solid blue circle in (a) is presented for V$_{\mathrm{G1}}=\unit[-750]{mV}$ in (b) for V$_{\mathrm{SD}}=\unit[-10]{mV}$ and (c) for V$_{\mathrm{SD}}=\unit[+10]{mV}$. While for negative V$_{\mathrm{SD}}$ current can freely flow through the base of the triangles, it is blocked for positive V$_{\mathrm{SD}}$ at zero magnetic field. PSB is lifted for a detuning $\epsilon\geq\epsilon_{\mathrm{ST}}$ or by applying a small magnetic field, here $B=\unit[40]{mT}$. The detuning axis is defined as indicated by the white arrow. A charge transport cycle is depicted schematically in (d) for negative and (e) for positive V$_{\mathrm{SD}}$.}
\label{fig2}
\end{figure*}
Integration of the self-aligned second gate layer follows after processing of the Si fin and the first gate level. These preceding fabrication steps are described in detail elsewhere\cite{Kuhlmann2018}. In Fig.\ \ref{fig1}\,(a) a scanning electron microscope (SEM) image of a device at this fabrication stage is shown. The first gate layer consists of a central nanoscale finger gate (G1) and two lead gates (L1\,\&\,L2) for source and drain reservoirs. The gaps separating these gates create channels (turquoise highlights in Fig.\ \ref{fig1}\,(a)) that will serve as a template for the second gate layer. By means of atomic layer deposition the gate stack, consisting of \unit[$\simeq$\,4.5]{nm} silicon oxide (SiO$_x$) and \unit[$\simeq$\,20]{nm} metallic titanium nitride (TiN), is deposited with highly uniform surface coverage (see Fig.\ \ref{fig1}\,(b)). Provided the width of the gaps separating G1 from L1/L2 is less than twice the thickness of the deposited material, the channels are almost evenly filled with metal. The thin SiO$_x$ layer ensures a good electrical isolation of the two gate layers (breakdown voltage \unit[$\gtrsim$\,6]{V}, see supporting information S1). 

Subsequently, an anisotropic TiN dry etch is applied for a duration that corresponds to the deposited metal thickness of \unit[$\simeq$\,20]{nm} (see Fig.\ \ref{fig1}\,(c)). While the gate metal is removed during etching from the flat surfaces of the device, leftovers are found at the topography steps. The TiN residues inside the predefined channels naturally form gates G2\,\&\,G3 of the second gate layer. The fan-out of the gates to microscale contact pads at larger distance from the fin is protected during etching by a resist mask (magenta highlights in Fig.\ \ref{fig1}\,(c)), which is defined by means of electron-beam lithography (EBL). The demands of this lithography step with regard to resolution and alignment accuracy are lowered by moving the channel endpoints further away from the center of the fin. 

At this stage, the gates of the second gate layer are still connected via the TiN that remains and is marked in red in Fig.\ \ref{fig1}\,(c\,\&\,d). This short circuit is eliminated by first protecting the gates with a resist mask, as shown in Fig.\ \ref{fig1}\,(d), and then by selectively removing all the unintentional TiN residues by isotropic wet etching (see Fig.\ \ref{fig1}\,(e)).  The protective cover is defined by means of EBL. 

After successful integration of the second gate layer, cross-sectional transmission electron microscope (TEM) images along and perpendicular to the fin direction are taken (see Fig.\ \ref{fig1}\,(f)). These images confirm ultra-small gate lengths as well as perfect gate alignment. All remaining fabrication steps, leading to a functional device, are described elsewhere\citep{Kuhlmann2018}.

The device layout with the three nanoscale gates G1, G2 and G3 allows for both a single- and double-dot operation mode. The latter one is explored in this work for $p$-type devices (see supporting information S2 for single-dot regime). Holes are accumulated in source (S) and drain (D) reservoirs through platinum silicide contacts, by a strong negative lead gate voltage (V$_{\mathrm{L1,\,L2}} = \unit[-4.5]{\mathrm{V}}$). Gates G2 and G3 form dots 1 and 2 and control their occupancy (see inset of Fig.\ \ref{fig2}\,(a) for a simplified equivalent circuit of the device). Gate G1, which is sandwiched between G2 and G3, is used to control the inter-dot tunnel coupling (see supporting information S3). The gate lengths of the device used here are \unit[$\simeq$\,25]{nm} for inter-dot barrier gate and \unit[$\simeq$\,15]{nm} for the plunger gates.

The data presented here are obtained from direct current electrical transport measurements with the sample cooled to \unit[0.55]{K}. In Fig.\ \ref{fig2}\,(a), a double dot charge stability diagram, showing the first observable pairs of bias triangles, is presented\cite{Wiel2002} (see supporting information S4 for same measurement on similar device). The two triangles of each pair strongly overlap for a source-drain voltage of V$_{\mathrm{SD}}=\unit[+10]{mV}$. While the lines of strong current flow parallel to the triangle base reveal elastic tunneling between the ground or excited states of the double dot, the background current inside the triangles can be assigned to inelastic tunneling\cite{Wiel2002}. The triangles for more negative gate voltages are distorted by co-tunneling processes because of the dots' increased tunnel coupling to the reservoirs\cite{Franceschi2001,Zumbuehl2004}. 

\begin{figure}[tb]
\centering \includegraphics[width=\linewidth]{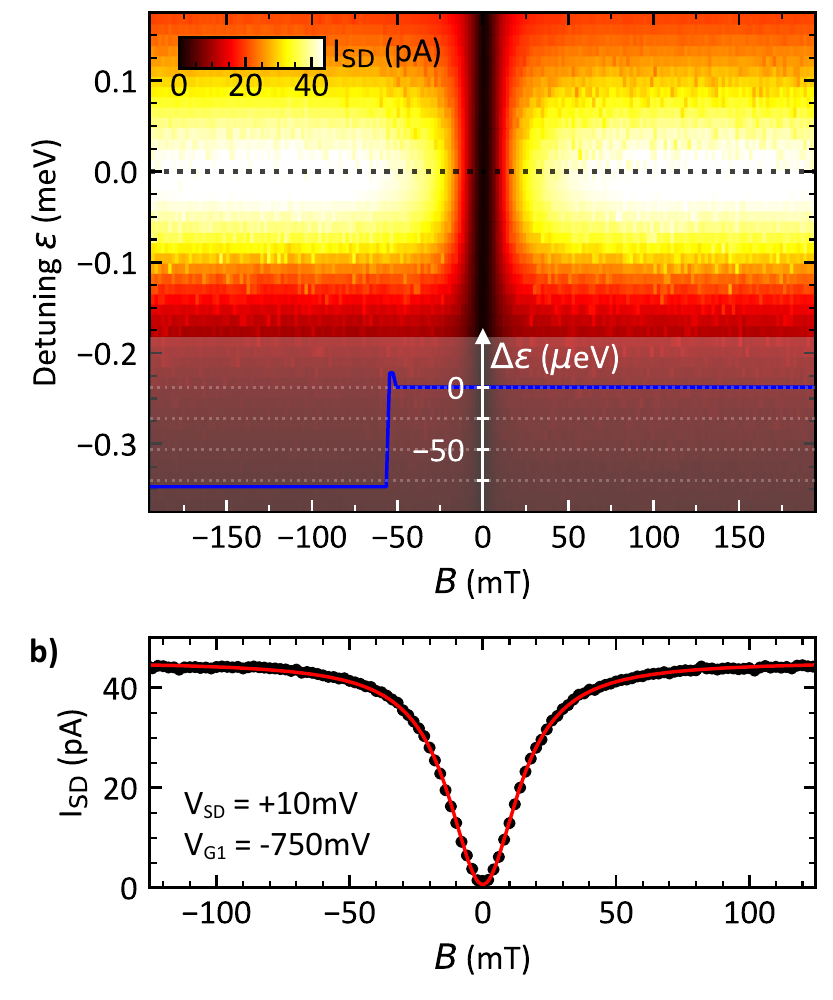}
\caption{\textbf{Spin blockade leakage current}. (a) Source-drain current I$_{\mathrm{SD}}$ under reverse bias as a function of detuning $\epsilon$ and out-of-plane magnetic field $\mathrm{B}$. Some of the vertical traces are shifted along the $\epsilon$-axis to eliminate the random switching of a charge trap. The detuning correction $\Delta\epsilon$ is plotted in the inset. (b) Cut along $\mathrm{B}$ at $\epsilon=0$, as indicated by the black dashed line in (a). The data (black dots) are well fitted by a Lorentzian function (red curve) of $\mathrm{FWHM}=\unit[32]{mT}$.}
\label{fig3}
\end{figure}

\begin{figure*}
\centering \includegraphics[width=\textwidth]{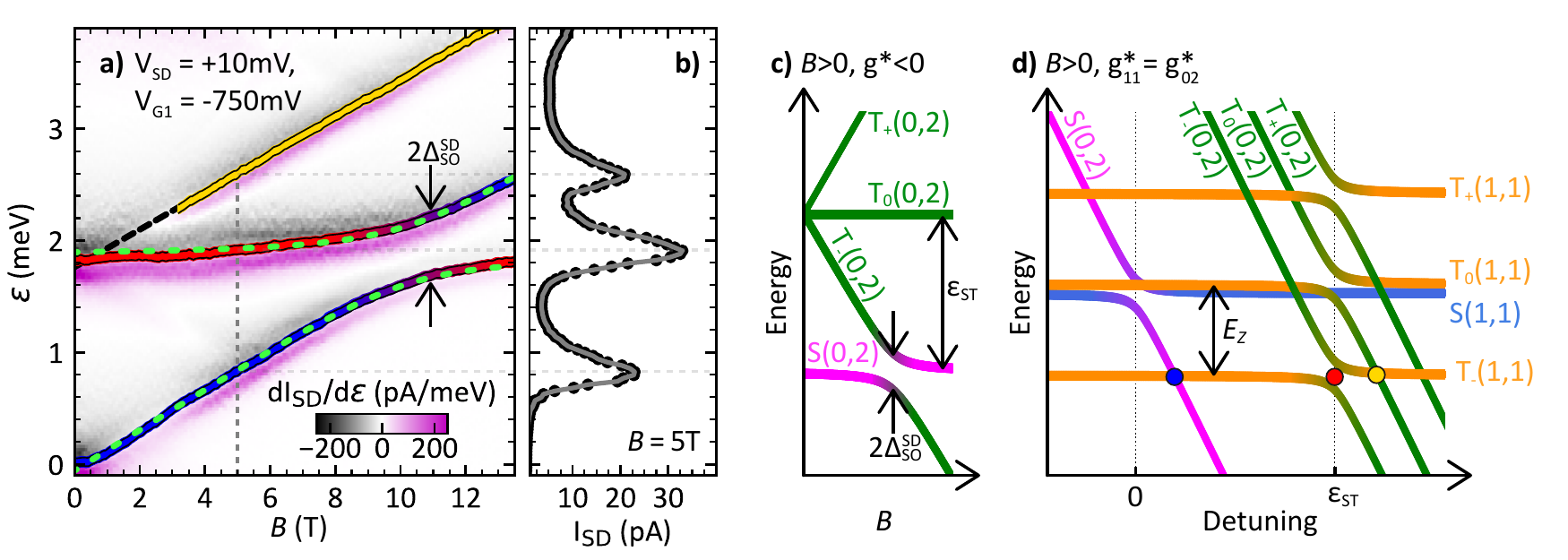}
\caption{\textbf{Spin state mixing by SOI}. (a) Detuning energy of the three observed resonant current peaks as a function of magnetic field. The peak positions are extracted from current traces, such as the one presented in (b) for $B=\unit[5]{T}$ (grey dashed line). The differentiated data dI$_{\mathrm{SD}}/\mathrm{d}\epsilon$ is shown in the background. The anticrossing of the two bottom curves is fitted with the standard expression for two-level repulsion (green dashed curves). For $B\unit[\lesssim3]{T}$ the energy splitting of the top and central peak cannot be resolved. (c) Magnetic field dependence of the double dot energy levels for $(0, 2)$ charge configuration. The degeneracy of the triplet spin states T$_+$, T$_0$ and T$_-$ is lifted by the Zeeman splitting $E_Z$. For $E_Z=\epsilon_{\mathrm{ST}}$ the S$(0, 2)$ and T$_{-}(0, 2)$ states hybridize due to SOI and anticross. (d) Double dot energy diagram versus detuning at finite magnetic field. Spin-conserved tunneling induces avoided crossings between states that share the same spin polarization\citep{Hanson2007}. The blue, red and yellow dots mark the transitions denoted with the same colors in (a).} 
\label{fig4}
\end{figure*}
In Figs.\ \ref{fig2}\,(b),\,(c) a zoom-in on the pair of bias triangles, indicated by a solid blue circle in Fig.\ \ref{fig2}\,(a), is presented for negative and positive V$_{\mathrm{SD}}$. While current flow through the base of the triangles is observed for V$_\mathrm{SD}=\unit[-10]{mV}$, it is strongly suppressed for positive V$_{\mathrm{SD}}$ at zero magnetic field $B$. (For the bias triangles marked by a solid magenta circle in Fig.\ \ref{fig2}\,(a) current suppression is observed for the opposite bias direction, see supporting information S5). This type of current rectification is a hallmark of PSB\cite{Ono2002,Johnson2005,Hanson2007,Lai2011,Li2015} and is due to spin-conserved tunneling, as schematically depicted in Figs.\ \ref{fig2}\,(d),\,(e). If two hole spins reside on the same dot (here the right one), they must occupy a spin singlet state S$(0,2)$ as the triplet state T$(0,2)$ is shifted to higher energy by the single-dot singlet-triplet splitting $\epsilon_{\mathrm{ST}}$\cite{Hanson2007}. Here $(m,n)$ denotes the effective hole occupancy of the left and right dot. While our data is consistent with observing the last hole, more sensitive charge detection methods are required to evaluate this\cite{Field1993,Hanson2007}. For a negative V$_\mathrm{SD}$ charge transport occurs from the S$(0,2)$ state to the S$(1,1)$ state, and a hole can escape the left dot to the reservoir. In contrast, for a positive V$_\mathrm{SD}$ current flow is blocked. If one hole spin resides on each dot, they can form either a S$(1,1)$ or T$(1,1)$ state, which are nearly degenerate in energy for weak inter-dot coupling. Once the T$(1,1)$ state is occupied by loading a hole from the reservoir to the left dot, transport is blocked by spin conservation during tunneling. 

PSB is lifted for an inter-dot energy level detuning $\epsilon$ exceeding $\epsilon_{\mathrm{ST}}$, since the T$(0,2)$ becomes accessible from the T$(1,1)$ state. Hence, the reappearance of current along the detuning axis determines $\epsilon_{\mathrm{ST}}\unit[\simeq1.85]{m\mathrm{eV}}$. This allows us to give an upper-bound estimate of the effective dot size $\lambda_x\sim\hbar/\sqrt{m^*\epsilon_{\mathrm{ST}}}=\unit[9.5]{nm}$\cite{Fasth2007}, which is in good agreement with the device geometry. Here, we assume harmonic confinement and an effective hole mass $m^*=0.45\,m_0$, where $m_0$ denotes the bare electron mass\cite{Kloeffel2018}.

For $\epsilon<\epsilon_{\mathrm{ST}}$ spin relaxation leads to a leakage current through the spin blocked region of the bias triangles\cite{Pfund2007,Churchill2009,Danon2009,NadjPerge2010}. For a small magnetic field of $B=\unit[40]{mT}$ current leaks through the base of the triangles (see Fig.\ \ref{fig2}\,(c)). For $B\neq0$ the previously forbidden T$(1, 1)\rightarrow \mathrm{S}(0,2)$ transition becomes allowed since hole spins in Si experience a strong SOI\cite{Kloeffel2018} that hybridizes the T$(1,1)$ and S$(0,2)$ states\citep{Danon2009,NadjPerge2010}.

The leakage current dependence on both $B$ and $\epsilon$ for positive V$_\mathrm{SD}$ is shown in Fig.\ \ref{fig3}\,(a). A dip in the leakage current, which is centered around zero magnetic field, is revealed by a line-cut along $B$ at $\epsilon=0$ in Fig.\ \ref{fig3}\,(b). This dip has a Lorentzian lineshape with a full-width-at-half-maximum (FWHM) of $\unit[32]{mT}$ and a close-to-zero minimum value, signifying a very efficient blockade. The dip also confirms that lifting of PSB is dominated by SOI\citep{Danon2009,Li2015}, since hyperfine interactions\cite{Pfund2007,NadjPerge2010} or spin-flip cotunneling \cite{Qassemi2009,Lai2011,Biesinger2015} yield a zero-field peak. 

An extension of Fig.\ \ref{fig3}\,(a) to both larger magnetic fields and detunings, revealing spin-orbit mediated singlet-triplet mixing, is presented in Fig.\ \ref{fig4}\,(a). Resonant charge transport occurs for detunings, where $(1, 1)$ and $(0, 2)$ spin states are degenerate in energy and hybridized by a finite coupling (see Fig.\ \ref{fig4}\,(d)). As seen in Fig.\ \ref{fig4}\,(b), three current peaks are observed within the $\epsilon$ range of Fig.\ \ref{fig4}\,(a), showing the $B$-dependence of the peak positions. For weak tunnel coupling and negative effective hole $g$-factor $g^*$, the bottom two curves in Fig.\ \ref{fig4}\, (a) can be assigned to the $\mathrm{T}_{-}(1, 1)\rightarrow\mathrm{S}(0, 2)$ and $\mathrm{T}_{-}(1, 1)\rightarrow\mathrm{T}_{-}(0, 2)$ transitions\cite{Li2015}. Thus, the $\mathrm{T}_{-}(1, 1)$ state can be used to probe the energy splitting of the S$(0, 2)$ and $\mathrm{T}_{-}(0, 2)$ states (see Fig.\ \ref{fig4}\,(c)).

While the central line remains at constant detuning for $B\unit[\lesssim3]{T}$, signifying a spin-conserving transition, the bottom line shifts by the Zeeman energy $E_Z=g_{11}^{*}\mu_{B}B$, where $g_{11}^{*}$ denotes the $g^{*}$-factor of the $(1, 1)$ triplet states and $\mu_{B}$ the Bohr magneton. From the slope we can thus extract $|g_{11}^{*}| = 3.2\pm0.3$, corresponding to an effective hole g-factor of $|g^{*}| = 1.6\pm0.2$, a value similar to those reported before\cite{Li2015,Voisin2015}. When $g_{02}^{*}\mu_{B}B$ approaches $\epsilon_{\mathrm{ST}}$, the S$(0,2)$ and T$_{-}(0, 2)$ states first begin to align in energy but then anticross due to SOI (see Fig.\ \ref{fig4}\,(c))\cite{Fasth2007}. This level repulsion causes the avoided crossing of the two bottom curves in Fig.\ \ref{fig4}\,(a) at $B_c\unit[\simeq10.9]{T}$. From the magnitude of the anticrossing we can extract the single-dot spin orbit gap $\Delta_{\mathrm{SO}}^{\mathrm{SD}}=\unit[0.27\pm0.03]{meV}$. The spin-orbit length can be estimated to be $\lambda_{\mathrm{SO}}\sim g^* \mu_B B_c \lambda_x/(\sqrt{2}\Delta_{\mathrm{SO}}^{\mathrm{SD}})= \unit[48]{nm}$ \cite{Fasth2007,Gao2020}, which is roughly half the value reported for holes in planar Si quantum dot structures\cite{Li2015}. Using $\epsilon_{\mathrm{ST}}=\unit[1.85]{meV}$ in addition to the parameters mentioned before we can overlay our data with the standard expression for two-level repulsion\cite{Fasth2007} (green curves in Fig.\ \ref{fig4}\,(a)) and find good agreement. 


In conclusion, we have introduced a novel self-alignment technique facilitating the fabrication of Si quantum devices with multiple gate layers. We employ such devices for reproducible formation of low-disorder double quantum dots and study spin-dependent hole transport. From the observation of spin blockade we extract a single-dot singlet-triplet splitting $\epsilon_{\mathrm{ST}}\unit[\simeq1.85]{m\mathrm{eV}}$, indicating large orbital energies due to ultra-small gate lengths. The magnetic field dependence of the leakage current identifies SOI as the dominant spin blockade lifting mechanism. An effective hole spin g-factor $|g^*|=1.6$ and single-dot spin orbit gap $\Delta_{\mathrm{SO}}^{\mathrm{SD}}= \unit[0.27]{meV}$ are derived by modelling a two-level anticrossing occurring at $B_c=\unit[10.9]{T}$. These results demonstrate that hole spins in Si are a promising candidate for building a scalable network of small, fast and electrically controllable qubits. 

Self-aligned gates have great potential for application in integrating spin-based multi-qubit devices. For holes in Si, qubit performance can further be enhanced by an optimized device geometry towards an ultra-strong and electrically switchable SOI\cite{Kloeffel2018, Froning2020}.

We thank C. Kloeffel and D. Loss for fruitful discussions, and U. Drechsler and A. Olziersky for technical support in device fabrication. This work was partially supported by the Swiss Nanoscience Institute (SNI), the NCCR QSIT, the Georg H. Endress Foundation, Swiss NSF (grant nr. 179024), and the EU H2020 European Microkelvin Platform EMP (grant nr. 824109).

\bibliography{fab_paper_refs}

\end{document}